\documentstyle[11pt]{article}
\begin{document}
\baselineskip 16pt

\vspace{1cm}
\centerline{\large\bf A Unified Treatment for XXX-Heisenberg Model and}
\centerline{\large\bf  Haldane-Shastry Model Using Shift Operators }
\vspace{1cm}

\centerline{ Jing-Ling Chen$^1$, Mo-Lin Ge$^2$, Kang Xue$^3$,
            and Xian-Geng Zhao$^1$}
\vspace{0.5cm}
{\small
\centerline{\bf 1. Laboratory of Computational Physics,}
\centerline{\bf Institute of Applied Physics and Computational Mathematics,}
\centerline{\bf P. O. Box 8009(26), Beijing 100088, P. R. China}
\centerline{ Email:   jinglingchen@eyou.com}
\vspace{0.4cm}
\centerline{\bf 2. Theoretical Physics Division, Nankai Institute of
            Mathematics,}
\centerline{\bf Nankai University, Tianjin 300071, P. R. China}
\centerline{ Email:   12geml@nankai.edu.cn}
\centerline{ Fax: 0086-22-23502408}
\vspace{0.4cm}
\centerline{\bf 3. Physics Department, Northeast Normal University,}
\centerline{\bf Changchun, Jilin, 130024, P. R. China}
}
\vspace{2cm}

\centerline{\bf Abstract}
\vspace{0.3cm}
 A unified treatment is developed for the XXX-Heisenberg model and a
 long-ranged interaction model (the $H_2$ in Haldane-Shastry model)
 from the point of view of shift operators (or raising and lowering
 operators), based on which the energy spectra of the spin-chain
 models are determined. Some physical discussions are also made.

\vspace{2cm}



\newpage


\centerline{\large\bf A Unified Treatment for XXX-Heisenberg Model and}
\centerline{\large\bf  Haldane-Shastry Model Using Shift Operators }

\vspace{1cm}

 \noindent
{\bf I. Introduction and Motivation}

\vspace{3mm}

 Recently much attention has been paid to long-ranged interaction
 models \cite{calogero}-\cite{vieira}.
 This is because such models are believed to be relevant to the quantum
 Hall effect and high $T_c$ superconductivity. The $S=1/2$
 Haldane-Shastry model (HSM) is one kind of typical and integrable
 long-ranged interaction models independently introduced by Haldane and
 Shastry \cite{haldane}\cite{shastry}. The following mutually commuting
 Hamiltonians are found for the HSM in \cite{haldane-1}\cite{wangzf}:

 $$
 H_{2}={\sum_{ij}}^{\prime} \biggr(
 \frac{z_iz_j}{z_{ij}z_{ji}} \biggr)(P_{ij}-1),
 $$
 $$
 H_{3}={\sum_{ijk}}^{\prime} \biggr(
 \frac{z_iz_jz_k}{z_{ij}z_{jk}z_{ki}} \biggr)(P_{ijk}-1),
 $$
 $$
 H_{4}={\sum_{ijkl}}^{\prime} \biggr(
 \frac{z_iz_jz_kz_l}{z_{ij}z_{jk}z_{kl}z_{li}}\biggr)(P_{ijkl}-1) +H'_4,
 $$
 \begin{equation}
    \label{N-hs-234}
 H'_4=-\frac{1}{3}H_2-2{\sum_{ij}}^{\prime} \biggr(
 \frac{z_iz_j}{z_{ij}z_{ji}}\biggr)^2(P_{ij}-1),
  \end{equation}
 where $P_{ij}$ is an operator that exchanges the spins at sites
  $i$ and $j$, $P_{ijk}$ and $P_{ijkl}$ are the cyclic permutation
operators:
 $$
 P_{ijk}=P_{ij}P_{jk} +P_{jk}P_{ki} +P_{ki}P_{ij},
 $$
 $$
 P_{ijkl}=P_{ij}P_{jk}P_{kl} + {\rm Cyclic} (i\rightarrow j\rightarrow
 k\rightarrow l).
 $$
 The primed sum omits equal values of the summation variables and
 $z_{ij}\equiv z_i-z_j$, $z_j= \exp(i2\pi j/N)$. For spin $S=1/2$,
 $P_{ij}= 1/2+2{\vec {\bf S}}_i\cdot{\vec {\bf S}}_j$. In this case,
 $H_2$, the first member of the Hamiltonian family of HSM reads
 ($H_{hs} \propto H_2$, and $N=$even for simplicity)

 \begin{equation}
   \label{N-hs}
 H_{hs}=4\sum_{i<j}^{N} J_{ij}
 \biggr({\vec {\bf S}}_i\cdot{\vec {\bf S}}_j- \frac{1}{4}\biggr), \;\;
 J_{ij}=J_0/\sin^2[(j-i)\pi/N],
 \end{equation}
 whose ground-state energy and correlation functions have been obtained
 together with the thermodynamics \cite{haldane}-\cite{gebhard}.
 The ground-state wave function is a spin singlet of the Jastrow-Gutzwiller
 form. The excitations are spin-$1/2$ spinons that form a gas of a
 semionic nature \cite{haldane}.
 Furthermore, the $H_{hs}$ is a natural extension of the XXX-Heisenberg
 model (XHM) \cite{mattis}$-$\cite{tennant} defined by

 \begin{equation}
 \label{Hxxx}
  H_{xxx}=J\sum_{j=1}^{N} \biggr({\vec {\bf S}}_j\cdot{\vec {\bf S}}_{j+1}
  -\frac{1}{4}\biggr),
 \end{equation}
 which is a nearest-neighbor interaction model exactly solved by the
 traditional Bethe ansatz \cite{bethe}.

 It can be observed that, eigenfunctions of spin-$1/2$ chain
 models (such as $H_{xxx}$ and $H_{hs}$ )
 with $r(r=0,1,2,\cdots,N)$ down-spins (i.e. with
 the eigenvalue $S_z=\frac{N}{2}-r$, where $S_z$ is the $z$-component of
 the total spin) have always the same form, they are

 \begin{equation}
 \label{N-eqpsir}
 \mid \psi_r>= \sum_{m_1<m_2<\cdots<m_r}
 a(m_1,m_2,\cdots,m_r)\phi( m_1,m_2,\cdots,m_r),
 \end{equation}
 where $\phi (m_1,m_2,\cdots,m_r)$ represents a spin state with $r$
 down-spins on the $m_j$-th $(j=1,2,\cdots,r)$ sites. In particular,
 $ \mid \psi_0>=\mid \uparrow \uparrow \cdots \uparrow > $
 is the vacuum state with all spins up of the spin-chain models.
 $a(m_1,m_2,\cdots,m_r)$ are coefficients defined for the ordering
 $m_1<m_2<\cdots<m_r$ and depend only upon $m_1, m_2, \cdots, m_r$.
 The determination of $a(m_1,m_2,\cdots,m_r)$ will be depended on
 particular model, i.e., XHM or HSM. The fact that eigenfunctions of
 different spin-chain models share the same form prompts us to develop
 a unified treatment for these models.

 Although the XHM and HSM have been studied for a long time,
 to our knowledge, there has not been a unified treatment yet. The
 purpose of this paper is to provide a unified treatment for the
 $H_{xxx}$ and $H_{hs}$ from the point of view of shift operators
 (or raising and lowering operators), based on which the energy spectra
 of the models are determined. The paper is organized as follows. In
 Sec. II, since the eigenfunctions of $H_{xxx}$ and $H_{hs}$ have the
 same form [see Eq. (\ref{N-eqpsir})], so we shall focus on
 Eq. (\ref{N-eqpsir}) and determine the forms of shift operators from
 physical considerations, and provide clear physical pictures how the
 raising operator $Q_{r,r-1}^+$ (or lowering operator $Q_{r-1,r}^-$)
 transforms $\mid \psi_{r-1}>$ to $\mid \psi_r>$ (or $\mid \psi_{r}>$ to
 $\mid \psi_{r-1}>$). In Sec. III and VI, we shall derive explicit results
 of shift operators for $H_{xxx}$ and $H_{hs}$ through the
 following definition of shift operators in a commutator form:

  \begin{equation}
  \label{N-eqHQ}
 (\;[H, Q_{r,r-1}^+] = \omega_{r,r-1} Q_{r,r-1}^+ \;)\; \mid \psi_{r-1}>,
\;\;
 (\;[H, Q_{r-1,r}^-] = \omega_{r-1,r} Q_{r-1,r}^- \;)\; \mid \psi_{r}>,
 \end{equation}
 where $ \omega_{nm} = E_n- E_m$ is the energy interval, $E_r$ is the energy
 corresponding to $\mid \psi_r>$. Some discussions are made in the last
 section.
 
 \vspace{3mm}
 \noindent
 {\bf II. Determining the Forms of Shift Operators from Physical
Consideration}

 \vspace{3mm}

 Let us focus on $\mid \psi_r>$ as shown in Eq. (\ref{N-eqpsir}), and
 assume that $a(m_1,m_2,\cdots,m_r)$ are unknown coefficients at all. We
 shall determine the forms of shift operators from physical considerations.
 Firstly, we consider the raising operator that satisfies

 \begin{equation}
 \label{N-eqqr2}
  Q_{r,r-1}^+
  \; \mid \psi_{r-1}>=\mid \psi_r>,
 \end{equation}
 namely, the raising operator $Q_{r,r-1}^+$ transforms the adjacent
 eigenstates specified by $r-1$ and $r$. An arbitrary state $\mid \psi_r>$
 can be obtained by repeated applications of $Q_{r,r-1}^+ $ to the given
 state $\mid \psi_0>$:
 $\mid \psi_r>= Q_{r,r-1}^+ Q_{r-1,r-2}^+ \cdots Q_{2,1}^+ Q_{1,0}^+
            \mid \psi_0>$.
 Taking Eq. (\ref{N-eqqr2}) and $H \mid \psi_{r}>=E_r \mid \psi_{r}>$ into
 account, one can easily verify Eq. (\ref{N-eqHQ}).

 Guided by the observation that
 $-i ({\vec S}_j \times {\vec S}_k)^- = S_j^- S_k^z - S_k^- S_j^z$
 and

 \begin{equation}
 \label{N-eqsjk2'}
 -i ({\vec S}_j \times {\vec S}_k)^- \mid \stackrel{j}{\uparrow}
 \stackrel{k}{\uparrow} > = \frac{1}{2}
 (\mid \stackrel{j}{\downarrow} \stackrel{k}{\uparrow} > -
 \mid \stackrel{j}{\uparrow} \stackrel{k}{\downarrow} > ),
 \;\;\;
 -i ({\vec S}_j \times {\vec S}_k)^- \mid \stackrel{j}{\downarrow}
 \stackrel{k}{\uparrow} > = \frac{1}{2}
 \mid \stackrel{j}{\downarrow} \stackrel{k}{\downarrow} > ,
 \end{equation}
 i.e. $-i ({\vec S}_j \times {\vec S}_k)$ make a transition between singlet
 and triplet states of ${\vec S}_j \cdot {\vec S}_k$,
 we should expect to obtain that, after introducing the following unified
 raising operator

 \begin{equation}
 \label{N-eqqr1}
  Q_{r,r-1}^+ = -i \sum_{j<k}^{N} W_{jk}^{(r)}({\vec S}_j \times {\vec
S}_k)^-,
 \;\; W_{jk}^{(r)}=-W_{kj}^{(r)},
 \end{equation}
 Eq. (\ref{N-eqqr2}) might be satisfied. Of course, other forms of raising
 operators such as
 $ {\cal Q}_{r,r-1}^+ =  \sum_{j=1}^{N} \alpha_j^{(r)} S_j^- $ can
 also be introduced. Direct calculations show that
 \begin{equation}
 \label{N-eqq1TOr}
    -i \sum_{j<k}^{N} W_{jk}^{(r)}({\vec S}_j \times {\vec S}_k)^-
  \; \mid \psi_{r-1}>= \sum_{j=1}^{N} \alpha_j^{(r)} S_j^- \mid \psi_{r-1}>
  =\; \mid \psi_{r}> ,
 \end{equation}
 where $ \alpha_j^{(r)}= \frac{1}{2} \sum_{k\neq j}^{N} W_{jk}^{(r)}$.
 In other words, when $Q_{r,r-1}^+$ acts on $\mid \psi_{r-1}>$, it can be
 simplified to
 ${\cal Q}_{r,r-1}^+ = \sum_{j=1}^{N} \alpha_j^{(r)} S_j^- $, whose
 form is more simple. However, $ \alpha_j^{(r)}$ is more complicate than
 $W_{jk}^{(r)}$, hence is much more difficult to determine than
$W_{jk}^{(r)}$.
 It is the reason why we introduce $Q_{r,r-1}^+$ in the beginning, but not
 ${\cal Q}_{r,r-1}^+$. Eq. (\ref{N-eqq1TOr}) means that ${\cal Q}_{r,r-1}^+$
 has the same effect as $Q_{r,r-1}^+$, they are both the raising operators.

 To find the forms of $Q_{r,r-1}^+$, we need to determine the forms of
 $W_{jk}^{(r)}$. In the following, we shall consider the cases
 $r=1,2,\cdots$ successively.

 {\bf (a)} $r=1$. In this case, Eq. (\ref{N-eqqr2}) becomes

 \begin{equation}
 \label{N-eqq0to1}
  \biggr[\; Q_{1,0}^+ = -i \sum_{j<k}^{N} W_{jk}^{(1)}({\vec S}_j \times
{\vec S}_k)^-
  \;\biggr] \mid \psi_{0}>=\mid \psi_1>.
 \end{equation}
 Due to $\mid \psi_1>=\sum_{m=1}^{N}a(m)\phi(m)$ and
 $ -i ({\vec S}_j \times {\vec S}_k)^-
 \mid \stackrel{j}{\uparrow}
 \stackrel{k}{\uparrow} > = \frac{1}{2}
 (\mid \stackrel{j}{\downarrow} \stackrel{k}{\uparrow} > -
 \mid \stackrel{j}{\uparrow} \stackrel{k}{\downarrow} > )$,
 by comparing the corresponding coefficients of $\phi(m)$ of the both sides
of
 Eq. ({\ref{N-eqq0to1}}), we have
 \begin{equation}
 \label{N-w1}
  \frac{1}{2}[W_{m1}^{(1)}+\cdots+W_{m,m-1}^{(1)}+W_{m,m+1}^{(1)}+
 \cdots+W_{mN}^{(1)}] =a(m),
 \;\;\; (m=1,2,\cdots).
 \end{equation}
 Under periodic boundary condition $a(m+N)=a(m)$, it is well known
 that $\sum_{m=1}^{N}a(m)= 0$, so the solutions of Eq. (\ref{N-w1}) are

 \begin{equation}
 \label{N-eqwjk1}
 W_{jk}^{(1)}=\frac{2}{N}(a(j)-a(k)), \;\; (j,k=1,2,\cdots,N).
 \end{equation}

 {\bf (b)} $r=2$. In this case, Eq. (\ref{N-eqqr2}) becomes
 \begin{equation}
 \label{N-eqq1to2}
  \biggr[\; Q_{2,1}^+ = -i \sum_{j<k}^{N} W_{jk}^{(2)}({\vec S}_j \times
{\vec S}_k)^-
  \;\biggr] \mid \psi_{1}>=\mid \psi_2>,
 \end{equation}
 where $ \mid \psi_2>=\sum_{m_1<m_2}^{N}a(m_1,m_2)\phi(m_1,m_2)$, and
 $\phi(m_1,m_2)=\mid \cdots \stackrel{m_1}{\downarrow} \cdots
 \stackrel{m_2}{\downarrow} \cdots >$
 represents a spin state with two down-spins on the $m_j$-th $(j=1,2)$
sites.
 After using $-i ({\vec S}_j \times {\vec S}_k)^-
 \mid \stackrel{j}{\downarrow} \stackrel{k}{\uparrow} > = \frac{1}{2}
 \mid \stackrel{j}{\downarrow} \stackrel{k}{\downarrow} >$
 and comparing the corresponding coefficients of $\phi(m_1,m_2)$ of the
 both sides of Eq. ({\ref{N-eqq1to2}}), we have

 \begin{equation}
 \label{N-briefw}
 \frac{1}{2}W_{m_1,m_2}^{(2)}[a(m_1)-a(m_2)]+
 \frac{1}{2}\sum_{m\ne m_1,m_2}^{N}[W_{m_2,m}^{(2)}a(m_1)+
    W_{m_1,m}^{(2)}a(m_2)]=a(m_1,m_2).
 \end{equation}
 with $ m_1,m_2=1,2,\cdots,N $.

 We now come to solve Eq. (\ref{N-briefw}), i.e. to find the solutions of
 $W_{m_1,m_2}^{(2)}$. Because $a(m)$ depends only on $m$,
 we have
 $$
 \biggr[\biggr(\frac{\partial a(m_1)}{\partial m_1}\biggr)^{-1}
   \frac{\partial}{\partial m_1}\biggr] a(m_1)=1, \;\;
 \biggr[\biggr(\frac{\partial a(m_1)}{\partial m_1}\biggr)^{-1}
   \frac{\partial}{\partial m_1}\biggr] a(m_2)=0,
 $$
 $$
 \biggr[\biggr(\frac{\partial a(m_2)}{\partial m_2}\biggr)^{-1}
   \frac{\partial}{\partial m_2}\biggr] a(m_2)=1, \;\;
 \biggr[\biggr(\frac{\partial a(m_2)}{\partial m_2}\biggr)^{-1}
  \frac{\partial}{\partial m_2}\biggr] a(m_1)=0,
 $$
 or in general

 \begin{equation}
 \label{N-partial2}
 \biggr[\biggr(\frac{\partial a(m)}{\partial m}\biggr)^{-1}
  \frac{\partial}{\partial m}\biggr] a(j)=\delta_{jm},
 \;\;\;(m,j=1,2,\cdots,N),
 \end{equation}
 where $m=x_m$ is the coordinate of the spin located on the $m-$th
 site of the lattice. From Eq. (\ref{N-partial2}), one can obtain
 \begin{equation}
 \label{N-select1}
 \biggr[\biggr(\frac{\partial a(m)}{\partial m}\biggr)^{-1}
  \frac{\partial}{\partial m}\biggr] \mid \psi_1>=
 \biggr[\biggr(\frac{\partial a(m)}{\partial m}\biggr)^{-1}
 \frac{\partial}{\partial m}\biggr]\sum_{m=1}^{N}a(m)\phi(m)
  =\phi(m),
 \end{equation}
 hence, the action of the partial differential operator $\biggr(
 \frac{\partial a(m)}{\partial m}\biggr)^{-1}\frac{\partial}{\partial m}$
 is clear, when it acts on $\mid \psi_1 >$, it will pick up the term
 $\phi(m)$ among $\mid \psi_1 >$.
           
 From the above analysis, if we set
 \begin{equation}
 \label{N-soluwjk}
 W_{j,k}^{(2)}=A(j,k)\;
 \biggr[ \biggr(\frac{\partial a(j)}{\partial j}\biggr)^{-1}
 \frac{\partial}{\partial j}- \biggr(\frac{\partial a(k)}{\partial k}
 \biggr)^{-1} \frac{\partial}{\partial k} \biggr],
 \end{equation}
 with
 \begin{equation}
\label{A2}
A(j,k)= A(k,j)= \left \{
  \begin{array}{ll}
     a(j,k)   &  {\rm if} \;\;j<k,  \\
     a(k,j)   &  {\rm if} \;\;j>k,
  \end{array}
   \right.
\end{equation}
 then Eq. (\ref{N-briefw}) is satisfied.

 One can note that $W_{j,k}^{(2)}$ shown in Eq. (\ref{N-soluwjk}) is a
partial
 differential operator acting on the coefficients $a(j)$.
 In the following, we shall derive the form of shift operator $Q_{2,1}^+$
 from physical consideration, so that physical picture transforming
 $\mid \psi_1>$ into $\mid \psi_2>$ can be seen clearly. In the former
 state, one of $N$ spins is down, while in the latter, two of $N$ spins are
 down. The crucial point is that, based on $\mid \psi_1>$ in which there has
 already been one down-spin, how we can invert the second one so that
 it can be shifted to $\mid \psi_2>$? Firstly, we must know clearly on which
 site the spin is down. The given state
 $\phi(m)= \mid \stackrel{1}{\uparrow} \stackrel{2}{\uparrow} \cdots
 \stackrel{m}{\downarrow} \cdots \stackrel{N}{\uparrow}  > $
 indicates that only on the $m$-th site the spin-state is down. With the
help
 of the partial differential operator $\frac{1}{\partial a(m)/\partial m}
 \frac{\partial}{\partial m}$, the wanted term can be
 picked up from $\mid \psi_1>$ [see Eq. (\ref{N-select1})]. After this
 manipulation, we know clearly that on the $m-$th site the spin has
 already been down, thus the second down-spin can occur on the $j$-th
 $ (j=1, 2, \cdots, m-1, m+1, \cdots, N)$ sites, respectively.
 Meanwhile, $\mid \psi_2 >$ can be rewritten as
 $$
 \mid \psi_2 >= \sum_{m_1<m_2}^{N}a(m_1,m_2)\phi(m_1,m_2)=
\frac{1}{2}  \sum_{m=1}^{N} \Phi(m),
$$
\begin{equation}
 \Phi(m)= \sum_{j=1}^{m-1}a(j,m)\phi(j,m) +
          \sum_{k=m+1}^{N}a(m,k)\phi(m,k).
\end{equation}
Introducing
$$
T_m=\biggr[ \; \sum_{j=1}^{m-1}T_{m \rightarrow (j,m)}^-  +
     \sum_{k=m+1}^{N}T_{m \rightarrow (m,k)}^- \;\biggr],
$$
since
$$
[ \; T_{m \rightarrow (j,m)}^-=
 a(j,m)(2i)({\vec S}_{j} \times {\vec S}_{m})^- \;]
\phi(m)= a(j,m)\phi(j,m),\;\;\;(j<m),
$$
$$
[ \; T_{m \rightarrow (m,k)}^-=
a(m,k)(-2i)({\vec S}_{m} \times {\vec S}_{k})^- \;]
\phi(m)= a(m,k)\phi(m,k),\;\;\;(k>m),
$$
one can obtain $T_m \phi(m) = \Phi(m)$. Define
$ F_m= T_m \biggr(\frac{\partial a(m)}{\partial m}\biggr)^{-1}
\frac{\partial}{\partial m}$, we then have
$$
F_m \mid \psi_1 >=
        \sum_{j=1}^{m-1}a(j,m)\phi(j,m) +
        \sum_{k=m+1}^{N}a(m,k)\phi(m,k)
$$
so that
\begin{equation}
\biggr(\sum_{m=1}^{N}F_m\biggr) \mid \psi_1 >= 2 \mid \psi_2 >.
\end{equation}
Since $
[\; Q_{2,1}^+ = -i \sum_{j<k}^{N} W_{jk}^{(2)}({\vec S}_j \times {\vec
S}_k)^-\;]
\mid \psi_1>=\mid \psi_2>$, one then arrives at
\begin{equation}
\label{N-eqqf}
Q_{2,1}^+ = \frac{1}{2}\sum_{m=1}^{N}F_m,
\end{equation}
by identifying the coefficients of $({\vec S}_j \times {\vec S}_k)^-$ of the
both sides of Eq. (\ref{N-eqqf}), it leads to
$$
W_{j,m}^{(2)}=a(j,m)\;
\biggr[ \biggr(\frac{\partial a(j)}{\partial j}\biggr)^{-1}\frac{\partial}
{\partial j} -  \biggr(\frac{\partial a(m)}{\partial m}\biggr)^{-1}
\frac{\partial}{\partial m}\biggr],
$$
\begin{equation}
W_{m,k}^{(2)}=a(m,k)\;
\biggr[ \biggr(\frac{\partial a(m)}{\partial m}\biggr)^{-1}
\frac{\partial}{\partial m} - \biggr(\frac{\partial a(k)}{\partial k}
\biggr)^{-1} \frac{\partial}{\partial k}\biggr],
\end{equation}
i.e. they are nothing but the explicit expression of Eq. (\ref{N-soluwjk}).

Therefore, the physical picture for the transformation from $\mid \psi_1>$
to $\mid \psi_2>$ is clear. This idea can be extended to generate
$\mid \psi_{r}>$ for given $\mid \psi_{r-1}>$, for example, in the  similar
manner we can find the shift operator transforming $\mid \psi_2>$ to
$\mid \psi_3>$.

{\bf (c)} $r=3$. Similar to {\bf (b)}, the eigenfunction
$\mid \psi_3>$ can be recast to
$\mid \psi_3>= \frac{1}{3} \sum_{m_1<m_2}^{N} \Phi(m_1,m_2)$,
where
$$
\Phi(m_1,m_2) =
     \sum_{j=1}^{m_1-1} a(j,m_1,m_2)\phi(j,m_1,m_2)
$$
$$
    + \sum_{k=m_1+1}^{m_2-1} a(m_1,k,m_2)\phi(m_1,k,m_2) +
     \sum_{l=m_2+1}^{N} a(m_1,m_2,l)\phi(m_1,m_2,l) .
$$
In $\mid \psi_2>$ two of $N$ spins are down, while in $\mid \psi_3>$ there
are three down-spins. Just like the case in {\bf (b)}, we introduce the
partial differential operator
$ \biggr(\frac{\partial^2 a(m_1,m_2)}{\partial m_1 \partial m_2}\biggr)^{-1}
 \frac{\partial}{\partial m_1}\frac{\partial}{\partial m_2},$
which yields
\begin{equation}
\label{N-selem1m2}
\biggr(\frac{\partial^2 a(m_1,m_2)}{\partial m_1 \partial m_2}\biggr)^{-1}
 \frac{\partial}{\partial m_1}\frac{\partial}{\partial m_2}
\mid \psi_2>= \phi(m_1,m_2),
\end{equation}
i.e. the term
$
\phi(m_1,m_2) = \mid \cdots \stackrel{m_1}{\downarrow} \cdots
\stackrel{m_2}{\downarrow}\cdots  >
$
is picked up from $\mid \psi_2>$, so that two down-spins on the $m_1-$th and
$m_2-$th sites have been pre-set. Further, the third down-spin can occur
on the $j-$th site:
$$
(j=1, 2, \cdots, m_1-1, m_1+1, m_1+2 ,\cdots, m_2-1,
m_2+1, m_2+2, \cdots, N).$$

Introducing
$$
T_{(m_1,m_2)}=
     \sum_{j=1}^{m_1-1} T_{(m_1,m_2) \rightarrow (j,m_1,m_2)}^-
$$
$$      +     \sum_{k=m_1+1}^{m_2-1} T_{(m_1,m_2) \rightarrow (m_1,k,m_2)}^-
     + \sum_{l=m_2+1}^{N} T_{(m_1,m_2) \rightarrow (m_1,m_2,l)}^-,
$$
where
$$
 T_{(m_1,m_2) \rightarrow (j,m_1,m_2)}^-=
 a(j,m_1,m_2) \;(2i) [\;
 ({\vec S}_{j} \times {\vec S}_{m_1})^-
 +
 ({\vec S}_{j} \times {\vec S}_{m_2})^- \;],
$$
$$
 T_{(m_1,m_2) \rightarrow (m_1,k,m_2)}^-=
 a(m_1,k,m_2) \;(2i) [\;
 -
 ({\vec S}_{m_1} \times {\vec S}_{k})^-
 +
 ({\vec S}_{k} \times {\vec S}_{m_2})^- \;],
$$
$$
 T_{(m_1,m_2) \rightarrow (m_1,m_2,l)}^-=
 a(m_1,m_2,l) \;(2i) [\;
 -
 ({\vec S}_{m_1} \times {\vec S}_{l})^-
 -
 ({\vec S}_{m_2} \times {\vec S}_{l})^- \;].
$$
One can verify that
$
T_{(m_1,m_2)} \phi(m_1,m_2) = \Phi(m_1,m_2).
$
Further, define
$$
F_{(m_1,m_2)}= T_{(m_1,m_2)}
   \biggr(\frac{\partial^2 a(m_1,m_2)}{\partial m_1 \partial
m_2}\biggr)^{-1}
 \frac{\partial}{\partial m_1}\frac{\partial}{\partial m_2},
$$
we get
$$
\biggr(\sum_{m_1<m_2}^{N}F_{(m_1,m_2)}\biggr) \mid \psi_2 >  =
       3 \mid \psi_3 >,
$$
on the other hand,
$ Q_{3,2}^+ \mid \psi_2>=\mid \psi_3>$,
thus
\begin{equation}
\label{N-eqq3f}
Q_{3,2}^+ = \frac{1}{3}\sum_{m_1<m_2}^{N}F_{(m_1,m_2)},
\end{equation}
by making comparison the coefficients of $({\vec S}_j \times {\vec S}_k)^-$
of the both sides of Eq. (\ref{N-eqq3f}), it yields
\begin{equation}
\label{N-soluw3jk}
W_{j,k}^{(3)}=  \frac{2}{3}
\biggr\{ \;
 \sum_{l\ne j,k}^{N} A(j,k,l)
 \biggr[ \biggr(\frac{\partial^2 A(j,l)}{\partial j \partial l }\biggr)^{-1}
     \frac{\partial}{\partial j} -
   \biggr(\frac{\partial^2 A(k,l)}{\partial k \partial l }\biggr)^{-1}
     \frac{\partial}{\partial k} \biggr]
   \frac{\partial}{\partial l}
\; \biggr\}.
\end{equation}
with
$$
A(j,k,l)= \left \{
  \begin{array}{ll}
     a(j,k,l)   &  {\rm if} \;\;j<k<l,  \\
     a(l,j,k)   &  {\rm if} \;\;l<j<k,   \\
     a(j,l,k)   &  {\rm if} \;\;j<l<k.   \\
  \end{array}
   \right.
$$
Thus the operator $Q_{3,2}^+$ shifting $\mid \psi_2>$ to $\mid \psi_3>$ is
also found.

Making use of the similar analysis, one can obtain the general form of
$Q_{r,r-1}^+$ with the solution of
$$
W_{j,k}^{(r)}  = \frac{2}{r}
\biggr\{ \;
 \sum_{l_1,l_2,\cdots,l_{r-2}\ne j,k}^{N}
    A(j,k,l_1,l_2,\cdots,l_{r-2}) \times
$$
\begin{equation}
\label{N-soluwrjk}
  \biggr[ \biggr(\frac{\partial^{r-1} A(j,l_1,l_2,\cdots,l_{r-2})}
 {\partial j \partial l_1 \partial l_2 \cdots \partial  l_{r-2}}\biggr)^{-1}
 \frac{\partial}{\partial j} -
   \biggr(\frac{ \partial^{r-1} A(k,l_1,l_2,\cdots,l_{r-2})}
 {\partial k \partial l_1 \partial l_2 \cdots \partial  l_{r-2}}\biggr)^{-1}
   \frac{\partial}{\partial k} \biggr]
   \frac{\partial}{\partial l_1}
   \frac{\partial}{\partial l_2} \cdots
   \frac{\partial}{\partial l_{r-2}}
 \biggr\},
\end{equation}
where $A(j,k,l_1,l_2,\cdots,l_{r-2})$ has the similar meaning
as $A(j,k)$ shown in Eq. (\ref{A2}).

In {\bf (b)} and {\bf (c)} one can see that the actions of the partial
differential operators $\biggr(\frac{\partial a(m)}{\partial m}\biggr)^{-1}
\frac{\partial}{\partial m}$ and $\biggr(\frac{\partial^2 a(m_1, m_2)}
{\partial m_1 \partial m_2 }\biggr)^{-1}\frac{\partial}{\partial m_1}
\frac{\partial}{\partial m_2}$ are picking up the terms $\phi(m)$ and
$\phi(m_1,m_2)$ from $\mid \psi_1>$ and $\mid \psi_2>$, respectively.
Since $\mid \psi_0>$ has only one single term, we need not to introduce some
partial differential operators to pick it up from $\mid \psi_0>$. That is
the reason why partial differential operators do not emerge in the
coefficients $W_{j,k}^{(1)}$. In the end of this section, we would like to
point out that (i) raising operator $Q_{1,0}^+ $ can be derived from
physical
consideration [see Appendix A]; (ii) The form of lowering operator
$Q_{r-1,r}^- $ can also are obtained following the same spirit as {\bf (b)}
and {\bf (c)} [see Appendix B].

\vspace{3mm}
 \noindent
{\bf III. Shift Operators for $H_{xxx}$ and Energy Spectrum}

\vspace{3mm}

To find the explicit result of $Q_{r,r-1}^+$ for $H_{xxx}$, we need to
determine the unknown coefficients $W_{jk}^{(r)}$ or
 $a(m_1,m_2,\cdots,m_r)$. The direct calculation shows
$$
[H_{xxx}, Q_{r,r-1}^+] = -J \sum\limits_{j,k=j+1}^{N} W_{j,j+1}^{(r)}
    \{ [({\vec S}_{j-1} \times {\vec S}_j) \times {\vec S}_{j+1}]^-
$$
$$
   + \frac{1}{2}( S_j^- - S_{j+1}^-)
     - [{\vec S}_{j} \times ({\vec S}_{j+1} \times {\vec S}_{j+2})]^-
     \}
 -J \sum\limits_{j,k \ge j+2}^{N} W_{j,k}^{(r)}
    \{ [({\vec S}_{j-1} \times {\vec S}_j) \times {\vec S}_{k}]^-
$$
\begin{equation}
\label{comut}
    - [({\vec S}_{j} \times {\vec S}_{j+1}) \times {\vec S}_{k}]^-
     + [{\vec S}_{j} \times ({\vec S}_{k-1} \times {\vec S}_{k})]^-
    - [{\vec S}_{j} \times ({\vec S}_{k} \times {\vec S}_{k+1})]^-
    \} .
\end{equation}
In the following, only $\mid \psi_0>$ is presumed known, its corresponding
energy is $E_0=0$. Now we consider the cases $r=1,2,\cdots$ successively.

(a) $r=1$. After acting Eq. (\ref{comut}) on $\mid \psi_0 >$, one obtains

$$
[H_{xxx}, Q_{1,0}^+] \mid \psi_0 >  =
-iJ \sum\limits_{j,k=j+1}^{N}
 \biggr[\biggr(\frac{W_{j+1,j+2}^{(1)}+W_{j-1,j}^{(1)}}
      {2W_{j,j+1}^{(1)}}-1 \biggr) W_{j,j+1}^{(1)}
     ({\vec S}_{j} \times {\vec S}_{j+1})^- \biggr] \mid \psi_0>
$$
$$
+ iJ \sum\limits_{j,k=j+1}^{N}[(W_{j+1,j+2}^{(1)}S_{j+2}^- +
     W_{j-1,j}^{(1)}S_{j-1}^-) ({\vec S}_{j} \times {\vec S}_{j+1})^z]
     \mid \psi_0>
$$
$$
- iJ \sum\limits_{j,k \ge j+2}^{N}
\biggr[ \biggr(\frac{W_{j+1,k+1}^{(1)}+
      W_{j-1,k-1}^{(1)}}{2W_{j,k}^{(1)}}-1 \biggr)W_{jk}^{(1)}
    ({\vec S}_{j} \times {\vec S}_k)^- \biggr] \mid \psi_0>
$$
$$
+  iJ \sum\limits_{j,k \ge j+2}^{N}W_{jk}^{(1)}
    [ S_{k}^- ({\vec S}_{j-1} \times {\vec S}_{j})^z
    -S_{k}^- ({\vec S}_{j} \times {\vec S}_{j+1})^z
$$
\begin{equation}
\label{r1}
 -     S_{j}^- ({\vec S}_{k-1} \times {\vec S}_{k})^z
    +S_{j}^- ({\vec S}_{k} \times {\vec S}_{k+1})^z  ]
     \mid \psi_0> .
\end{equation}
Since $ ({\vec S}_{j} \times {\vec S}_{k})^z \mid \psi_0 >=0 $,
Eq. (\ref{r1}) then becomes
\begin{equation}
\label{r1-1}
[H_{xxx}, Q_{1,0}^+] \mid \psi_0 >  =
 -iJ \sum\limits_{j<k}^{N} \biggr[ \biggr(\frac{W_{j+1,k+1}^{(1)}+
      W_{j-1,k-1}^{(1)}}{2W_{jk}^{(1)}}-1 \biggr)W_{jk}^{(1)}
    ({\vec S}_{j} \times {\vec S}_k)^- \biggr] \mid \psi_0>.
\end{equation}
Comparing Eq. (\ref{r1-1}) with the first of Eq. (\ref{N-eqHQ}),
$Q_{1,0}^+$ is a raising operator unless the factor
$[W_{j+1,k+1}^{(1)}+ W_{j-1,k-1}^{(1)}]/2W_{jk}^{(1)}$
is a real number and does not depend on $j$ and $k$. If set

\begin{equation}
\label{w-1}
      W_{jk}^{(1)} =
      \frac{2}{N} [a(j)-a(k)], \;\; a(j)=\exp(ij\theta),
\end{equation}
Eq. (\ref{r1-1}) yields
$[H_{xxx}, Q_{1,0}^+] \mid \psi_0 >  =  (\omega_{10} Q_{1,0}^+) \mid \psi_0
>
$, where

\begin{equation}
\label{omega1}
\omega_{10}= J  \biggr(\frac{W_{j+1,k+1}^{(1)}+
      W_{j-1,k-1}^{(1)}}{2W_{jk}^{(1)}}-1 \biggr)= J(\cos\theta -1)
\end{equation}
is the energy interval between $E_1$ and $E_0$. Under periodic boundary
condition $a(m+N)=a(m)$, it is well known that
$$
\theta =\frac{2\pi}{N}n;\ \ n= \pm 1,\cdots,\pm (\frac{N}{2}-1),
\pm \frac{N}{2},
$$
and $ \sum_{m=1}^{N}a(m)= 0$. Since $E_0=0$, Eq. (\ref{omega1}) leads
to $E_1= J(\cos\theta -1)$. After acting $Q_{1,0}^+$ on $\mid \psi_0 >$,
the next wavefunction $\mid \psi_1 >$ is obtained.

(b) $r=2$. We set
$\mid \psi_2>= \sum_{j<k}a(j,k)\phi( j,k)$
with unknown expansion coefficients $a(j,k)$. The direct calculation shows

$$
[H_{xxx}, Q_{2,1}^+] \mid \psi_1 >  =
J \sum\limits_{j,k=j+1}^{N} \biggr[\frac{a(j-1,j+1)+a(j,j+2)}
      {2a(j,j+1)}-2 \biggr] a(j,j+1) \phi(j,j+1)
$$
$$
+  J \sum\limits_{j,k \ge j+2}^{N}
\biggr[\frac{a(j-1,k)+a(j+1,k)+a(j,k-1)+a(j,k+1)}
      {2a(j,k)}-2\biggr] a(j,k) \phi(j,k)
$$

\begin{equation}
\label{r2}
- E_1 (Q_{2,1}^+ \mid \psi_1 >)
-  iJ \sum\limits_{j<k}^{N}
  \biggr\{
  \biggr[ \sum\limits_{j=1}^{N}
    ({\vec {\bf S}}_j\cdot{\vec {\bf S}}_{j+1}-\frac{1}{4})\biggr]
    ({\vec S}_{j} \times {\vec S}_k)^-
    W_{j,k}^{(2)} \sum\limits_{m \ne j,k}^{N} a(m) \phi (m) \biggr\}.
\end{equation}
To make $Q_{2,1}^+$ be a raising operator of $H_{xxx}$, we should require
\begin{equation}
\label{requ0}
 Q_{2,1}^+ \mid \psi_1 > = \mid \psi_2 > = \sum\limits_{j<k}a(j,k)\phi( j,k)
,
\end{equation}
\begin{equation}
\label{requ1}
 W_{jk}^{(2)} \sum\limits_{m \ne j,k}^{N} a(m) \phi (m)=0 ,
\end{equation}
and
\begin{equation}
\label{requ2}
  \begin{array}{lll}
 E_2 & =  & \frac{a(j-1,k)+a(j+1,k)+a(j,k-1)+a(j,k+1)} {2a(j,k)}-2  \\
     & =  &  \frac{a(j-1,j+1)+a(j,j+2)} {2a(j,j+1)}-2 .
   \end{array}
\end{equation}
Since $a(m)=\exp (im\theta)$ is an exponential function, then one can find
that
\begin{equation}
\label{partial2}
\frac{1}{i\theta}\frac{\partial}{\partial m} a(j)=\delta_{jm}a(m),
\;\;\;(m,j=1,2,\cdots,N),
\end{equation}
here $m=x_m$ is understood as the coordinate of the spin located on the
$m$-th site of the lattice.

Due to Eq. (\ref{partial2}), to make Eq. (\ref{requ1}) be valid, one finds
$ W_{jk}^{(2)}$ can be the following solution
\begin{equation}
\label{wjk2}
W_{jk}^{(2)}=  \frac{1}{i\theta} a(j,k)\;
\biggr[ \frac{1}{a(j)} \frac{\partial}{\partial (m=j)}-
  \frac{1}{a(k)} \frac{\partial}{\partial (m=k)} \biggr],
  \;\;\; (j<k)
\end{equation}
with a still unknown coefficient $a(j,k)$, which will be determined
 by requiring that the two factor of the right-hand side of Eq.
(\ref{requ2})
are real numbers and do not depend on $j$ and $k$. Obviously, if we choose
$a(j,k)$ to be the usual Bethe ansatz, i.e.,
\begin{equation}
\label{ajk}
a(j,k)=Ce^{ij\theta_1}e^{ik\theta_2}+C'e^{ij\theta_2}e^{ik\theta_1},\;\;
(j,k=1,2,\cdots,N)
\end{equation}
where $C$ and $C'$ are independent upon $j$ and $k$. Eq. (\ref{ajk}) yields
$$
\frac{a(j-1,k)+a(j+1,k)+a(j,k-1)+a(j,k+1)} {2a(j,k)}
=\cos \theta_1 +\cos \theta_2
$$
for arbitrary $C$ and $C'$. From Eq. (\ref{requ2}) we should require
\begin{equation}
\label{ansatz}
\frac{a(j-1,j+1)+a(j,j+2)} {2a(j,j+1)}
=\cos \theta_1 +\cos \theta_2 ,
\end{equation}
thus
\begin{equation}
\label{cc1}
\frac{C}{C'}=-\frac{1-2e^{i\theta_1}+e^{i(\theta_1+\theta_2)}}
{1-2e^{i\theta_2}+e^{i(\theta_1+\theta_2)}}.
\end{equation}
On the other hand, from the periodic condition $a(j,k)=a(k,j+N)$, we obtain
\begin{equation}
\label{cc2}
e^{iN\theta_1}=\frac{C}{C'}=e^{-iN\theta_2}.
\end{equation}

Actually, Eqs.(\ref{cc1}) and (\ref{cc2}) can be rewritten in the following
form

\begin{equation}
 \label{xx-eq46}
e^{iN\theta_1}=-e^{-i\Theta(\theta_1,\theta_2)},\ \
e^{iN\theta_2}=-e^{-i\Theta(\theta_2,\theta_1)},
\end{equation}
with
\begin{equation}
  \label{xx-eq47}
\Theta(\theta,\theta')=2 \arctan
\biggr\{ \frac{\sin \frac{\theta-\theta'}{2}}
{\cos \frac{\theta+\theta'}{2}-\cos \frac{\theta-\theta'}{2}} \biggr\}
=2\arctan \biggr\{\frac {1}{2}
\biggr[\cot \frac{\theta}{2}-\cot \frac{\theta'}{2}\biggr] \biggr\}
\end{equation}
is an odd function, i.e., $\Theta(\theta,\theta')=-\Theta(\theta',\theta)$.
From equation (\ref{xx-eq46}) we have
\begin{equation}
 \label{xx-eq48}
N\theta_1=2\pi I-\Theta(\theta_1,\theta_2),\ \ \
N\theta_2=2\pi I'-\Theta(\theta_2,\theta_1),
\end{equation}
where $I$ and $I'\; (I'\ne I)$ belong to the set $\{ \pm \frac{1}{2},
\pm \frac{3}{2},\cdots,\pm \frac{N-1}{2} \}$.

Thus we obtain two equations, i.e., (\ref{cc1}) and (\ref{cc2}),
for the parameters
$\theta_1$ and $\theta_2$, which determine the energy $E_2$:

\begin{equation}
\label{Energy2}
E_2=J(\cos \theta_1 +\cos \theta_2 -2)=
J\sum_{i=1}^{r=2}(\cos \theta_i -1).
\end{equation}
 Eq. (\ref{xx-eq48}) is nothing but the usual Bethe Ansatz equation,
 obviously shift operators recover the results obtained through
 the usual Bethe Ansatz method.

 If we select $C=\exp(i\frac{\phi_{12}}{2}),\; \;
C'=\exp(-i\frac{\phi_{12}}{2})$, from Eq. (\ref{xx-eq47}) we have

\begin{equation}
 \label{xx-eq49}
2\cot{\frac{\phi_{12}}{2}}=\cot \frac{\theta_1}{2}-\cot \frac{\theta_2}{2},
\hspace{1cm} -\pi \le \phi_{12}  \le \pi.
\end{equation}
After denoting $\phi_{21}=-\phi_{12},\ m=m_1,\ k=m_2 $, one obtains

\begin{equation}
 \label{xx-eq50}
a(m_1,m_2)=e^{i(\theta_1m_1+\theta_2m_2+\frac{\phi_{12}}{2})}+
e^{(i\theta_1m_2+\theta_2m_1+\frac{\phi_{21}}{2})},
\end{equation}
whose form is quite convenient to be generalized to the general case.
Moreover, since $a(j,k)$ is defined in the ordering $j<k$,
to make $W_{jk}^{(2)}=-W_{kj}^{(2)}$, we rewrite $W_{jk}^{(2)}$ as

\begin{equation}
\label{soluwjk}
W_{jk}^{(2)}=  \frac{1}{i\theta}
A(j,k)\;
\biggr[ \frac{1}{a(j)} \frac{\partial}{\partial (m=j)}-
  \frac{1}{a(k)} \frac{\partial}{\partial (m=k)} \biggr],
\end{equation}
where $A(j,k)$ is shown in Eq. (\ref{A2}).

(c) For general $r\ge 3$, This case is similar with case (b) by setting
\begin{equation}
 \label{xx-eq51}
a(m_1,m_2,\cdots,m_r)=\sum_{P=1}^{r!}\exp\biggr[i\biggr(\sum_{k=1}^{r}
{\theta_{P_k}}m_k+\frac{1}{2}\sum\phi_{P_k,P_n}\biggr)\biggr].
\end{equation}
Substituting $Q_{r,r-1}^+$ with
$$
W_{j,k}^{(r)}  = \frac{2}{r}
\frac{1}{i\theta_1}\frac{1}{i\theta_2}\cdots \frac{1}{i\theta_{r-1}}
\biggr\{ \;
 \sum_{l_1,l_2,\cdots,l_{r-2}\ne j,k}^{N}
    A(j,k,l_1,l_2,\cdots,l_{r-2})
$$
\begin{equation}
\label{soluwrjk}
 \biggr[\; \frac{1}{A(j,l_1,l_2,\cdots,l_{r-2})} \frac{\partial}{\partial j}
 - \frac{1}{A(k,l_1,l_2,\cdots,l_{r-2})} \frac{\partial}{\partial k} \;
 \biggr] \;
   \frac{\partial}{\partial l_1}
   \frac{\partial}{\partial l_2} \cdots
   \frac{\partial}{\partial l_{r-2}}
\; \biggr\},
\end{equation}
 into the first of Eq. (\ref{N-eqHQ}), and due to the periodic condition
$
a(m_2,\cdots,m_r,m_1+N)=a(m_1,m_2,\cdots,m_r)$,
 we then have
\begin{equation}
 \label{xx-eq54}
E_r=J\sum_{m=1}^{r}(\cos \theta_m-1),
\end{equation}
where $\theta_m$ satisfies the Bethe equations
$$
e^{iN\theta_m}=(-1)^{r-1}\prod\limits_{k=1}^{r}\exp[-i\Theta(\theta_m,
\theta_k)],
$$
or
\begin{equation}
 \label{xx-eq56}
N\theta_m=2\pi I_m-\sum\limits_{k=1}^{r} \Theta(\theta_m,\theta_k),
\hspace{4mm} (m\neq k)
\end{equation}
for $m=1,2,\cdots,r$, and $I_m$ runs over the $r$ distinct integers
$\{0,\pm 1,\cdots,\pm \frac{N}{2} \}$, for odd $r$, and $\{ \pm \frac{1}{2},
\pm \frac{3}{2},\cdots,\pm \frac{N-1}{2} \}$, for even $r$.

Lowering operator $Q_{r-1,r}^-$ can also be obtained by the similar way.
Further more, it is worth mentioning that, like the case discussed by
Bethe Ansatz method, the values of $\theta_m$ must be all distinct
because, otherwise, $a(m_1,m_2,\cdots,m_r)=0$ and $\mid \psi_r>=0$.
This is easily seen from equations (\ref{ajk}) and (\ref{xx-eq51})
for $r=2$.

\vspace{3mm}
 \noindent
{\bf IV. Shift Operators for $H_{hs}$ and Energy Spectrum}

\vspace{3mm}

To find the explicit result of $Q_{r,r-1}^+$ for $H_{hs}$, we need to
determine the corresponding unknown coefficient $W_{jk}^{(r)}$, or
$a(m_1,m_2,\cdots,m_r)$. The direct calculation shows

$$
[H_{hs}, Q_{r,r-1}^+] = 4 \sum\limits_{j<k}^{N} W_{j,k}^{(r)}
    \biggr\{ \sum\limits_{n=1,n\ne k-j}^{N-1}J_n
    [({\vec S}_{j} \times {\vec S}_{j+n}) \times {\vec S}_{k}]^-
$$
$$
  + \sum\limits_{n=1,n\ne N-(k-j)}^{N-1}J_n
    [{\vec S}_{j} \times ({\vec S}_{k} \times {\vec S}_{k+n})]^-
   - \frac{1}{2} J_{k-j}( S_j^- - S_{k}^-)
     \biggr\}
$$
$$
=  4 \sum\limits_{j<k}^{N} W_{j,k}^{(r)}
    \biggr\{ \sum\limits_{n=1,n\ne k-j}^{N-1} iJ_n
    [({\vec S}_{j} \times {\vec S}_{j+n})^- { S}_{k}^z-
    ({\vec S}_{j} \times {\vec S}_{j+n})^z { S}_{k}^-]
$$
\begin{equation}
\label{hs-comut}
   + \sum\limits_{n=1,n\ne N-(k-j)}^{N-1} iJ_n
    [ \; { S}_{j}^- ({\vec S}_{k} \times {\vec S}_{k+n})^z-
    { S}_{j}^z ({\vec S}_{k} \times {\vec S}_{k+n})^-]
 - \frac{1}{2} J_{k-j}( S_j^- - S_{k}^-)
     \biggr\}.
\end{equation}
where $J_n=J_{j,j+n}=J_0/\sin^2(n\pi/N)$, $J_{k-j}=J_0/\sin^2[(k-j)\pi/N]$.
In the following, only $\mid \psi_0>$ is presumed known, its corresponding
energy is $E_0=0$. Now we consider the cases $r=1,2,\cdots$ successively.

(a) $r=1$. After acting Eq. (\ref{hs-comut}) on $\mid \psi_0 >$, and
 using $ ({\vec S}_{j} \times {\vec S}_{k})^z \mid \psi_0 >=0 $, one then
 obtains
$$
[H_{hs}, Q_{1,0}^+] \mid \psi_0 >  =
-i \sum\limits_{j<k}^{N}\biggr[\biggr( \sum\limits_{n=1}^{N-1}
\frac{W_{j-n,k-n}^{(1)}} {W_{j,k}^{(1)}} 2J_n -
\sum\limits_{n=1}^{N-1} 2J_n \biggr) \;
 W_{j,k}^{(1)} ({\vec S}_{j} \times {\vec S}_{k})^-\biggr] \mid \psi_0>
$$
\begin{equation}
\label{hs-r1}
 - [W_{k-N,k}^{(1)} J_{k-j}\phi(j)- W_{j,j}^{(1)} J_{k-j}\phi(k)].
\end{equation}
Comparing Eq. (\ref{hs-r1}) with the first of Eq. (\ref{N-eqHQ}),
$Q_{1,0}^+$ is a raising operator unless the factor
$W_{j-n,k-n}^{(1)}/ W_{j,k}^{(1)}$
is a real number which does not depend on $j$ and $k$, and

\begin{equation}
\label{hs-w1=0}
      W_{k-N,k}^{(1)} = W_{j,j}^{(1)}=0.
\end{equation}
If we set
$     W_{jk}^{(1)} = 2[a(j)-a(k)]/N$,
then Eq. (\ref{hs-w1=0}) is satisfied, and

\begin{equation}
\label{hs-w1rq-1}
\frac{W_{j-n,k-n}^{(1)}} {W_{j,k}^{(1)}}=
\frac{a(j-n)-a(k-n)}{a(j)-a(k)},
\end{equation}
which is required to be a real number and independent on $j$ and $k$.
Taking the periodic boundary condition $a(m+N)=a(m)$ into account, we
have

\begin{equation}
\label{hs-w1rq-2}
   a(m)=e^{im\pi},
\end{equation}
so that $W_{j-n,k-n}^{(1)}/ W_{j,k}^{(1)}=e^{-in\pi}=(-1)^n$.

Denote
$ x=\sum\limits_{n=1}^{N-1}J_n=\frac{J_0}{3}(N^2-1), \;\;
  y=\sum\limits_{n=1}^{N-1}(-1)^{n+1} J_n=\frac{J_0}{3}(\frac{N^2}{2}+1)$,
Eq. (\ref{hs-r1}) yields
$[H_{hs}, Q_{1,0}^+] \mid \psi_0 >  =  (\omega_{10} Q_{1,0}^+)
\mid \psi_0 >$, where
\begin{equation}
\label{hs-omega1}
\omega_{10}= 2\biggr( \sum\limits_{n=1}^{N-1}(-1)^n J_n -
   \sum\limits_{n=1}^{N-1}J_n\biggr)=-2(x+y)
\end{equation}
is the energy interval between $E_1$ and $E_0$. Since $E_0=0$,
then $E_1= -2(x+y)$. After acting $Q_{1,0}^+$ on $\mid \psi_0 >$,
the next wave function $\mid \psi_1 >$ is obtained.

(b) $r=2$. We set
$\mid \psi_2>= \sum_{j<k}a(j,k)\phi( j,k)$
with unknown expansion coefficients $a(j,k)$. One can have

$$
[H_{hs}, Q_{2,1}^+] \mid \psi_1 >  =
 \sum\limits_{j<k}^{N}\biggr[ \sum\limits_{n=1}^{N-1} 2J_n
   \frac{a(j-n,k)+a(j,k-n)}{a(j,k)} -
   4 \sum\limits_{n=1}^{N-1}J_n
$$
$$
  +  4J_{k-j} -
   2J_{k-j}\frac{a(k-N,k)+a(j,j)}{a(j,k)}\biggr]a(j,k)\phi(j,k)
- E_1 (Q_{2,1}^+ \mid \psi_1 >)
$$
\begin{equation}
\label{hs-r2}
 - i \sum\limits_{j<k}^{N}
  \biggr\{
  \biggr[ \sum\limits_{j=1}^{N}
    4J_{k-j}\biggr({\vec {\bf S}}_j\cdot{\vec {\bf S}}_{k}-\frac{1}{4}
    \biggr)\biggr] ({\vec S}_{j} \times {\vec S}_k)^-
    W_{j,k}^{(2)} \sum\limits_{m \ne j,k}^{N} a(m) \phi (m) \biggr\}.
\end{equation}
To make $Q_{2,1}^+$ be a raising operator of $H_{hs}$, we should require
\begin{equation}
\label{hs-requ0}
 Q_{2,1}^+ \mid \psi_1 > = \mid \psi_2 > = \sum\limits_{j<k}a(j,k)\phi( j,k)
,
\end{equation}
\begin{equation}
\label{hs-requ1}
 W_{jk}^{(2)} \sum\limits_{m \ne j,k}^{N} a(m) \phi (m)=0 ,
\end{equation}
and
\begin{equation}
\label{hs-requ2}
  \begin{array}{lll}
 E_2  & = &
  \sum\limits_{n=1}^{N-1} 2J_n
   \frac{a(j-n,k)+a(j,k-n)}{a(j,k)} -
   4\sum\limits_{n=1}^{N-1}J_n \\
   && + 4J_{k-j} -
   2J_{k-j}\frac{a(k-N,k)+a(j,j)}{a(j,k)}.
   \end{array}
\end{equation}

To make Eqs.(\ref{hs-requ0}) and (\ref{hs-requ1}) be valid, one can set
$$
W_{j,k}^{(2)}=a(j,k)\;
\biggr[ \biggr(\frac{\partial a(j)}{\partial j}\biggr)^{-1}
\frac{\partial}{\partial j} -\biggr(\frac{\partial a(k)}{\partial k}
\biggr)^{-1} \frac{\partial}{\partial k}\biggr],
  \;\;\; (j<k)
$$
with a still unknown coefficient $a(j,k)$, which will be determined
 by requiring that the right-hand side of Eq. (\ref{hs-requ2})
be a real number that does not depend on $j$ and $k$.

 On one hand, since $\phi(m_1, m_2)$ represents a spin state with two
 down-spins on the $m_i$-th $(i=1,2)$ sites of the lattice, and there
 is only one spin on each site, $\phi(j,j)$ and $\phi(k,k)$ would not
 have physical meanings, hence we should require

 \begin{equation}
\label{hs-req-1}
a(j,j)=0, \;\;\; a(k-N,k)=a(k,k)=0.
\end{equation}
 On the other hand, $-4\sum_{n=1}^{N-1}J_n=-4x$ does not depend
 on $j$ and $k$, from Eq. (\ref{hs-requ2}) one requires

\begin{equation}
\label{hs-req-2}
 E'_2   =  \sum\limits_{n=1}^{N-1} 2J_n
   \frac{a(j-n,k)+a(j,k-n)}{a(j,k)}  + 4J_{k-j}
\end{equation}
be a real number that does not depend on $j$ and $k$. Now let us come to
find the coefficients $a(j,k)$ satisfying the requirements
Eqs.(\ref{hs-req-1}) and (\ref{hs-req-2}).

 {\bf (i)} Like the case in the XXX-Heisenberg model, in the beginning,
 we set the Ansatz

\begin{equation}
\label{hs-ajk}
a(j,k)=Ce^{ij\theta_1}e^{ik\theta_2}+C'e^{ij\theta_2}e^{ik\theta_1},\;\;
(j,k=1,2,\cdots,N),
\end{equation}
where $C$ and $C'$ are independent upon $j$ and $k$. Eq. (\ref{hs-ajk})
yields
\begin{equation}
\label{hs-ans1}
\frac{a(j-n,k)+a(j,k-n)}{a(j,k)}
=e^{-in \theta_1}+e^{-in \theta_2}
\end{equation}
for arbitrary $C$ and $C'$. After substituting Eq. (\ref{hs-ans1}) into
the $E'_2$ as shown in Eq. (\ref{hs-req-2}), one will find that the first
term
of $E'_2$, i.e.,
$\sum\limits_{n=1}^{N-1} 2J_n[a(j-n,k)+a(j,k-n)]/a(j,k)$
is independent upon $j$ and $k$, while the second term of $E'_2$, i.e.,
$4J_{k-j}$ depends on $j$ and $k$, so the Ansatz shown in Eq. (\ref{hs-ajk})
does not satisfy the requirement (\ref{hs-req-2}). Based on this
observation,
the factor $[a(j-n,k)+a(j,k-n)]/a(j,k)$ will have to depend on $j$ and $k$,
so that the summation
$\sum\limits_{n=1}^{N-1} 2J_n[a(j-n,k)+a(j,k-n)]/a(j,k)$ can contribute
a term $-4J_{k-j}$ to cancel the second term of $E'_2$.

 {\bf (ii)} We modify the Ansatz as following:

\begin{equation}
\label{hs-ajk-2}
a(j,k)=C[e^{ij\theta_1}e^{ik\theta_2}+\lambda]
+C'[e^{ij\theta_2}e^{ik\theta_1}+\lambda'],
\end{equation}
where $C$, $C'$, $\lambda$ and $\lambda'$ do not depend on $j$ and $k$.
Eq. (\ref{hs-ajk-2}) yields

\begin{equation}
\label{hs-ans2}
\frac{a(j-n,k)+a(j,k-n)}{a(j,k)}
=e^{-in \theta_1}+e^{-in \theta_2}+
\frac{2(C\lambda + C'\lambda')[1-(e^{-in \theta_1}+e^{-in \theta_2})/2]}
{C[e^{ij\theta_1}e^{ik\theta_2}+\lambda]
+C'[e^{ij\theta_2}e^{ik\theta_1}+\lambda']}.
\end{equation}

From Eq. (\ref{hs-ajk-2}) one obtains

\begin{equation}
\label{hs-ajj-1}
a(j,j)=C[e^{ij(\theta_1+\theta_2)}+\lambda]
+C'[e^{ij(\theta_1+\theta_2)}+\lambda']=
 (C+C')e^{ij(\theta_1+\theta_2)} +(C\lambda+C'\lambda'),
\end{equation}
since $a(j,j)=0$ for arbitrary $j$, we then have the following two kinds of
possibilities:

{\bf (ii-1)}: $C=-C', \;\; C\lambda+C'\lambda'=0$.

However, $C\lambda+C'\lambda'=0$ will let the factor
$[a(j-n,k)+a(j,k-n)]/a(j,k)$ do not depend on $j$ and $k$, so we eliminate
this possibility.

{\bf (ii-2)}:
\begin{equation}
\label{hs-theta12-1}
\lambda=\lambda'=-1, \;\;
\theta_1+\theta_2=0, \;\; {\rm or} \;\; 2\pi \times {\rm integer},
\end{equation}
so that

$$
a(j,k)=C[e^{-i(k-j)\theta_1}-1]+C'[e^{i(k-j)\theta_1}-1],
$$
\begin{equation}
\label{hs-ans2-2}
\frac{a(j-n,k)+a(j,k-n)}{a(j,k)}
=2\cos(n\theta_1)+
\frac{-2(C + C')[1- \cos(n\theta_1)]}{a(j,k)}.
\end{equation}
In order to cancel the second term of $E'_2$ (i.e., $4J_{k-j}$), we
should require

\begin{equation}
\label{hs-a(jk)-3}
a(j,k) \;\; \propto \;\; \frac{1}{4J_{k-j}} \;\; \propto \;\;
\sin^2[(k-j)\pi/N],
\end{equation}
or
\begin{equation}
\label{hs-a(jk)-4}
 \frac{C}{C'}[e^{-i(k-j)\theta_1}-1]+[e^{i(k-j)\theta_1}-1]
 \;\; \propto \;\; e^{i2(k-j)\pi/N} +e^{-i2(k-j)\pi/N} -2.
\end{equation}
On the other hand, the periodic condition $a(j,k)=a(k,j+N)$ yields
\begin{equation}
\label{hs-cc2}
e^{iN\theta_1}=\frac{C}{C'}=e^{-iN\theta_2}.
\end{equation}
thus one obtains

\begin{equation}
\label{hs-cc2-1}
C=C', \;\; \theta_1=-\theta_2=2\pi/N,
\end{equation}
which leads to

$$
\frac{a(j-n,k)+a(j,k-n)}{a(j,k)}
=2 \biggr[\; 1-\sin^2(n\pi/N)+
\frac{\sin^2(n\pi/N)}{\sin^2[(k-j)\pi/N]} \;\biggr].
$$
The summation
$$
\sum\limits_{n=1}^{N-1} 4J_n \sin^2(n\pi/N) /\sin^2[(k-j)\pi/N]=
\sum\limits_{n=1}^{N-1} 4J_0 /\sin^2[(k-j)\pi/N]=
(N-1)4J_{k-j}
$$
cannot cancel the second term of $E'_2$ (i.e. $4J_{k-j}$). However,
it is easy to observe that the summation
$\sum\limits_{n=1}^{N-1} (-1)^n 4J_0 /\sin^2[(k-j)\pi/N]=-4J_{k-j}$ can
reach the purpose. This observation renders us to modify again the Ansatz
as following:

 {\bf (iii)} We set the Ansatz:

\begin{equation}
\label{hs-a(jk)-2-1}
a(j,k)=e^{i\pi(j+k)} \{ C[e^{ij\theta_1}e^{ik\theta_2}+\lambda]
+C'[e^{ij\theta_2}e^{ik\theta_1}+\lambda'] \},
\end{equation}
or

\begin{equation}
\label{hs-ajk'}
a(j,k)= e^{i\pi (j+k)} \sin^2[(k-j)\pi/N],
\;\;
(j,k=1,2,\cdots,N)
\end{equation}
whose form is quite convenient to be generalized to the general case
for $r>2$. Eq. (\ref{hs-ajk'}) yields directly $a(k-N,k)=a(j,j)=0$, and

\begin{equation}
\label{hs-ajk-1}
 \frac{a(j-n,k)+a(j,k-n)}{a(j,k)}=
2 e^{-i\pi n} \biggr[1-2\sin^2(n\pi/N) +\frac{\sin^2(n\pi/N)}
{\sin^2[(k-j)\pi/N]}\biggr].
\end{equation}
From Eq. (\ref{hs-requ2}) we obtain

\begin{equation}
\label{hs-E2}
E_2= -4(x+y)+8J_0.
\end{equation}
After acting $Q_{2,1}^+$ on $\mid \psi_1 >$,
the next wave function $\mid \psi_2 >$ is obtained.

(c) $r=3$. We set

\begin{equation}
\label{hs-a3}
\mid \psi_3>= \sum_{m_1<m_2<m_3}a(m_1,m_2,m_3)\phi(m_1,m_2,m_3)
\end{equation}
with unknown expansion coefficients $a(m_1,m_2,m_3)$. The direct
calculation shows
$$
[H_{hs}, Q_{3,2}^+] \mid \psi_2 >  =
$$
$$
 \sum\limits_{m_1<m_2<m_3}^{N} \biggr\{ \sum\limits_{n=1}^{N-1} 2J_n
   \frac{a(m_1-n,m_2,m_3)+a(m_1,m_2-n,m_3)+a(m_1,m_2,m_3-n)}
   {a(m_1,m_2,m_3)}
$$
$$
 -  6\sum\limits_{n=1}^{N-1}J_n
    + 4(J_{m_2-m_1} + J_{m_3-m_1} + J_{m_3-m_2})
$$
$$
 -  \frac{2J_{m_2-m_1}[a(m_1,m_1,m_3)+a(m_2-N,m_2,m_3)]}{a(m_1,m_2,m_3)}
$$
$$
 -  \frac{ 2J_{m_3-m_1}[a(m_1,m_2,m_1)+a(m_3-N,m_2,m_3)]}{a(m_1,m_2,m_3)}
$$
$$
 -  \frac{ 2J_{m_3-m_2}[a(m_1,m_2,m_2)+a(m_1,m_3-N,m_3)]}{a(m_1,m_2,m_3)}
   \biggr\}
   a(m_1,m_2,m_3)\phi(m_1,m_2,m_3) 
$$
$$
     -E_2 (Q_{3,2}^+ \mid \psi_2 >)
$$
\begin{equation}
\label{hs-r3}
  -i \sum\limits_{j<k}^{N}
  \biggr\{
  \biggr[ \sum\limits_{j=1}^{N}
    4J_{k-j}\biggr({\vec {\bf S}}_j\cdot{\vec {\bf S}}_{k}-\frac{1}{4}
    \biggr)\biggr]  ({\vec S}_{j} \times {\vec S}_k)^-
    W_{j,k}^{(3)} \sum\limits_{m_1, m_2 \ne j,k}^{N}
    a(m_1,m_2) \phi (m_1,m_2) \biggr\}.
\end{equation}
To make $Q_{3,2}^+$ be a raising operator of $H_{hs}$, we should require
\begin{equation}
\label{hs-requ0-r3}
 Q_{3,2}^+ \mid \psi_2 > = \mid \psi_3 > =
 \sum_{m_1<m_2<m_3}a(m_1,m_2,m_3)\phi(m_1,m_2,m_3),
\end{equation}
\begin{equation}
\label{hs-requ1-r3}
    W_{j,k}^{(3)} \sum\limits_{m_1, m_2 \ne j,k}^{N}
    a(m_1,m_2) \phi (m_1,m_2)=0
\end{equation}
and
$$
E_3=\sum\limits_{n=1}^{N-1} 2J_n
   \frac{a(m_1-n,m_2,m_3)+a(m_1,m_2-n,m_3)+a(m_1,m_2,m_3-n)}
   {a(m_1,m_2,m_3)}
$$
$$
 -  6\sum\limits_{n=1}^{N-1}J_n
    + 4(J_{m_2-m_1} + J_{m_3-m_1} + J_{m_3-m_2})
$$
$$
 -  \frac{2J_{m_2-m_1}[a(m_1,m_1,m_3)+a(m_2-N,m_2,m_3)]}{a(m_1,m_2,m_3)}
$$
$$
 -  \frac{ 2J_{m_3-m_1}[a(m_1,m_2,m_1)+a(m_3-N,m_2,m_3)]}{a(m_1,m_2,m_3)}
$$

\begin{equation}
\label{hs-requ2-r3}
 -  \frac{ 2J_{m_3-m_2}[a(m_1,m_2,m_2)+a(m_1,m_3-N,m_3)]}{a(m_1,m_2,m_3)}
\end{equation}

To make Eqs.(\ref{hs-requ0-r3}) and (\ref{hs-requ1-r3}) be valid, one can
set
$$
W_{j,k}^{(3)}=  \frac{2}{3}
\biggr\{ \;
 \sum_{l\ne j,k}^{N} A(j,k,l)
 \biggr[ \biggr(\frac{\partial^2 A(j,l)}{\partial j \partial l }\biggr)^{-1}
     \frac{\partial}{\partial j} -
   \biggr(\frac{\partial^2 A(k,l)}{\partial k \partial l }\biggr)^{-1}
     \frac{\partial}{\partial k} \biggr]
   \frac{\partial}{\partial l}
\; \biggr\}.
$$
with a still unknown coefficient $a(j,k,l)$, which will be determined
 by requiring that the right-hand side of Eq. (\ref{hs-requ2-r3})
be a real number that does not depend on $j$ and $k$. If we choose
\begin{equation}
\label{hs-ajk-2'}
a(m_1,m_2,m_3)= e^{i \pi \sum_{i=1}^{3} m_i} \prod_{i<j}
\sin^2(\pi(m_j-m_i)/N),\;\;
(m_1,m_2,m_3=1,2,\cdots,N)
\end{equation}
it yields
$$
a(m_1,m_1,m_3)=a(m_2-N,m_2,m_3)=a(m_3-N,m_2,m_3)=0, \;\; etc.
$$
Denote
$$
  \lambda_1=(m_3-m_2)\pi/N, \;\;   \lambda_2=(m_2-m_1)\pi/N, \;\;
    \lambda_3=(m_3-m_1)\pi/N=\lambda_1+\lambda_2,
$$
we then have
$$
  \frac{a(m_1-n,m_2,m_3)+a(m_1,m_2-n,m_3)+a(m_1,m_2,m_3-n)}
   {a(m_1,m_2,m_3)}=
  e^{-in\pi} \biggr\{
$$
$$
  3[1-2\sin^2(n\pi/N)]^2 + 2[1-2\sin^2(n\pi/N)]\sin^2(n\pi/N)
      \sum\limits_{i=1}^{3} \biggr(\frac{1}{\sin^2(\lambda_i)}\biggr)
$$
$$
 + \sin^4(n\pi/N)
 \biggr(\frac{1}{\sin^2(\lambda_1)\sin^2(\lambda_2)}
 +\frac{1}{\sin^2(\lambda_1)\sin^2(\lambda_3)}
 +\frac{1}{\sin^2(\lambda_2)\sin^2(\lambda_3)} \biggr)
$$

\begin{equation}
\label{hs-ajk-3}
 + \sin^2(2n\pi/N) [ \cot(\lambda_1)\cot(\lambda_3)
 +\cot(\lambda_2)\cot(\lambda_3) -
 \cot(\lambda_1)\cot(\lambda_2) ] \}.
\end{equation}

Since
$$
 \cot(\lambda_1)\cot(\lambda_3) +\cot(\lambda_2)\cot(\lambda_3) -
 \cot(\lambda_1)\cot(\lambda_2)=-1,
$$
\begin{equation}
\label{hs-ajk-4}
 \sum\limits_{n=1}^{N-1}e^{-in\pi} \sin^2(n\pi/N)=0,\;\;
 \sum\limits_{n=1}^{N-1}e^{-in\pi} \cos^2(n\pi/N)=-1,\;\;
 \sum\limits_{n=1}^{N-1}e^{-in\pi}=-1,\;\;
\end{equation}
from Eq. (\ref{hs-requ2-r3}) one obtains

\begin{equation}
\label{hs-E3}
E_3= -6(x+y)+32J_0.
\end{equation}

 For general $r\geq 4$, by setting

$$
\mid \psi_r>= \sum_{m_1<m_2<\cdots<m_r}
a(m_1,m_2,\cdots,m_r)\phi( m_1,m_2,\cdots,m_r),
$$
\begin{equation}
  \label{hhs-a2-rr}
a(m_1,m_2,\cdots,m_r)= e^{i \pi \sum_{j=1}^{r} m_j}
\prod_{i<j} \sin^2(\pi(m_j-m_i)/N),
\end{equation}
and making use of the similar analysis, one can obtain shift operators
$Q_{r,r-1}^+$ and $Q_{r-1,r}^-$, and the energy $E_r$ corresponding to
$\mid \psi_r>$ is

\begin{equation}
\label{hs-Er}
E_r= -2r(x+y)+4r(r-1)J_0+\frac{4}{3}r(r-1)(r-2)J_0.
\end{equation}

\vspace{3mm}
 \noindent
{\bf V. Discussion and Conclusion}

\vspace{3mm}

A standard quantum mechanical transition problem in general has the
following format. The first quantity we must have is a Hamiltonian
${\cal H}$, which can be divided as $ {\cal H}=H_0 + H_I$,
where for some region of coordinate space or time $H_I$ can be neglected.
Secondly, when $H_I$ is neglected, it is meaningful of speak of the energy
levels and corresponding states of the free Hamiltonian $H_0$ between which
the transitions take place. These transitions are induced by the
interaction $H_I$. Experiments can detect the frequencies (i.e. the energy
intervals) satisfying the Bohr frequency condition $\omega_{nm} = E_n-
E_m$. The energy spectrum of $H_0$ can be determined from experiments
is owing to the existence of the external interaction $H_I$. The physical
nature of the external factor, which causes the quantum transition of the
microparticles is arbitrary. In particular, it may be the interaction of
the microparticles with electromagnetic radiation. Typical examples can
be seen in a hydrogen atom or a harmonic oscillator, where a transition
from one stationary state to another is realized by an electric dipole
moment. The dipole moment operator ${\hat {\bf d}}$ of any atom is
expressible as a sum of raising and lowering operators
${\hat {\cal L}}(n,m)$ between states $\mid \psi_m >$ and $\mid \psi_n >$
\cite{shore}: ${\hat {\bf d}} = \sum_{n,m}{\bf d}_{nm}{\hat {\cal L}}(n,m)$,
where
$$
{\hat {\cal L}}^+(n,m)= Q_{n,n-1}^+ Q_{n-1,n-2}^+ \cdots Q_{m+2,m+1}^+
   Q_{m+1,m}^+,
$$
$$
{\hat {\cal L}}^-(m,n)= Q_{m,m+1}^- Q_{m+1,m+2}^- \cdots Q_{n-2,n-1}^-
   Q_{n-1,n}^-, \;\;\;
 (n>m).
$$
Usually, in a hydrogen atom or a harmonic oscillator, the dipole moment
operator is the coordinate {\bf r} or the momentum {\bf p} of the particle,
and the interaction $H_I$ is expressed by the scalar product of the dipole
moment operator and the external field. We would like to extend this kind
of dipole transition mechanism to the XHM or HSM so that its energy spectrum
might be detected from experiments.

Fistly, we take the $H_{xxx}$ or $H_{hs}$ as the free Hamiltonian $H_0$.
Secondly, we write the interaction as
$H_I(t)= {\hat {\bf d}}\cdot {\bf B}(t)$,
where
${\bf B}(t)=\sum_{\lambda}[\;
{\bf e}(\lambda) {\cal E}_\lambda^{(+)}(t)+
{\bf e}^*(\lambda) {\cal E}_\lambda^{(-)}(t) \;]$
is a time-dependent magnetic field, and
$$
{\cal E}_\lambda^{(+)}(t)={\cal E}_\lambda(t) \exp(-i\omega_\lambda t),
\;\;\;
{\cal E}_\lambda^{(-)}(t)={\cal E}^*_\lambda(t) \exp(+i\omega_\lambda t).
$$
Hence, the time-dependent Schr{\" o}dinger equation is
$$
i\frac{\partial}{\partial t} \mid \Psi(t)>=
(H_0+{\hat {\bf d}}\cdot {\bf B}(t) ) \mid \Psi(t)>.
$$
The general state $\mid \Psi(t)>$ is written as an expansion
$$
\mid \Psi(t)>= \sum_{n=1} C_n(t)\mid \psi_n > \exp(-iE_n t),
$$
where $\mid \psi_n >$'s are eigenstates of $H_0$. We thereby obtain a set
couple equations
$$
  \begin{array}{lll}
-i\frac{d}{dt} C_n(t) &= &
\sum\limits_k \sum\limits_\lambda {\bf d}_{nk}
\cdot {\bf e}(\lambda) {\cal E}_\lambda
\exp[-i(\omega_\lambda -\omega_{nk})t] C_k(t)\\
& &  + \sum\limits_k \sum\limits_\lambda {\bf d}_{nk}
\cdot {\bf e}^*(\lambda) {\cal E}^*_\lambda
\exp[+i(\omega_\lambda -\omega_{nk})t] C_k(t),
  \end{array}
$$
with $\omega_{nk}=E_n-E_k$ is the Bohr transition frequency.
Obviously, when $\omega_\lambda =\omega_{nk}$, the magnetic resonance
phenomena would happen, thus the energy intervals of the spin chain
might be detected.

    Starting from the ferromagnetic ``vacuum" state $\mid \psi_0 >$,
if the raising operators are acted for enough times, then for even-spins
antiferromagnetic XHM (or HSM), it will reach the ground state. The
corresponding ground state energy was first calculated by Hulth{\'{e}}n
using Bethe's method \cite{hulthen}. The ground state is a singlet with
total spin $S_T=0$ (for $N=$even integer), therefore the number of
spin-deviates in
the ground state is $r=N/2$ (The proof that the total spin is indeed minimal
in the ground state is found in \cite{Lieb-M}). des Cloiseaux and Pearson
(dCP) were the first to study the elementary excitations \cite{desC}, which
they interpreted as spin-wavelike states with $S_T=1$. It was later shown by
Faddeev and Takhtajan \cite{Faddeev-3} that the natural excitations
(spinons)
actually have $S_T=1/2$, and hence fermions. The underlying excitations
occur only in pairs \cite{Faddeev-3}. The dCP states are now understood to
be a superposition of two spinons, one of which carries zero momentum.
If the energy spectrum is detected by the dipole transition mechanism,
people might ask: What are the first and the second excited states? And
what are their degeneracies? These problems are still open and under
investigation. In conclusion, we have developed a unified treatment for the
XXX-Heisenberg model and the $H_2$ in Haldane-Shastry model using
shift operators, based on which the energy spectra of the models are
determined. Furthermore, it is also interesting and significant to extend
the shift operator approach to the Hubbard model \cite{liebwu}\cite{korepin}
and the generalized Bethe ansatz \cite{yang} in subsequent investigations.


\vspace{1cm}

{\bf Acknowledgment}

This work was partially supported by the National Natural Science Foundation
of China.

\pagebreak

\vspace{3mm}
\noindent
{\bf APPENDIX A: Physical Picture of Shifting $\mid \psi_0>$ to
$\mid \psi_1>$}
\vspace{3mm}

Because
\begin{equation}
-i ({\vec S}_j \times {\vec S}_k)^- \mid \stackrel{j}{\uparrow}
\stackrel{k}{\uparrow} > = \frac{1}{2}
(\mid \stackrel{j}{\downarrow} \stackrel{k}{\uparrow} > -
\mid \stackrel{j}{\uparrow} \stackrel{k}{\downarrow} > ),
\;\;\;
(S_j^-+S_k^-)\mid \stackrel{j}{\uparrow} \stackrel{k}{\uparrow} >
=
(\mid \stackrel{j}{\downarrow} \stackrel{k}{\uparrow} > +
\mid \stackrel{j}{\uparrow} \stackrel{k}{\downarrow} > ),
\end{equation}
so that
$$
\mid \stackrel{j}{\downarrow} \stackrel{k}{\uparrow} >
=
[-i ({\vec S}_j \times {\vec S}_k)^-   +
\frac{1}{2}(S_j^-+S_k^-)]
\mid \stackrel{j}{\uparrow} \stackrel{k}{\uparrow} >,
$$
\begin{equation}
\label{N-spin}
\mid \stackrel{j}{\uparrow} \stackrel{k}{\downarrow} >
=
[i ({\vec S}_j \times {\vec S}_k)^-   +
\frac{1}{2}(S_j^-+S_k^-)]
\mid \stackrel{j}{\uparrow} \stackrel{k}{\uparrow} >.
\end{equation}
Due to Eq. (\ref{N-spin}), one finds the transformation
$$
\mid \stackrel{1}{\uparrow} \stackrel{2}{\uparrow} \cdots
\stackrel{m}{\uparrow} \cdots \stackrel{N}{\uparrow}  >
  \Longrightarrow
a(m)\phi(m)=
a(m) \mid \stackrel{1}{\uparrow} \stackrel{2}{\uparrow} \cdots
\stackrel{m}{\downarrow} \cdots \stackrel{N}{\uparrow}  >
\;\;\;(m=1,2,\cdots,N),
$$
can be realized by the operator:
\begin{equation}
 T_{0 \rightarrow m}^-= \frac{a(m)}{N-1} \biggr\{ \;
     \sum_{j=1}^{m-1} \biggr[
     i ({\vec S}_j \times {\vec S}_m)^-  + \frac{1}{2}(S_j^-+S_m^-) \biggr]
   +
     \sum_{j=m+1}^{N} \biggr[
    -i ({\vec S}_m \times {\vec S}_j)^-  + \frac{1}{2}(S_m^-+S_j^-) \biggr]
    \biggr\}.
\end{equation}
Define
$$
 F_0=\sum_{m=1}^{N} T_{0 \rightarrow m}^- =
$$
\begin{equation}
 \label{N-eqfpsi}
 -i \frac{1}{N-1}\sum_{j<k}^{N}[a(j)-a(k)] ({\vec S}_j \times {\vec S}_k)^-
+
 \frac{1}{N-1}\sum_{m=1}^{N}\biggr[\frac{N-2}{2}a(m)+
 \frac{1}{2} \sum_{k=1}^{N}a(k) \biggr]S_m^-,
\end{equation}
one gets
\begin{equation}
  \label{N-eqfpsin}
 F_0 \mid \psi_0 >
=\sum_{m=1}^{N} a(m)\phi(m) = \mid \psi_1 >
\end{equation}

Using
\begin{equation}
  \sum_{m=1}^{N} a(m)=0, \;\;\;
 \sum_{m=1}^{N}a(m)S_m^- \mid \psi_0 >=\mid \psi_1 >,
\end{equation}
from Eq. (\ref{N-eqfpsi}) and  Eq. (\ref{N-eqfpsin}) we then have
\begin{equation}
 -i \frac{1}{N-1}\sum_{j<k}^{N}[a(j)-a(k)] ({\vec S}_j \times {\vec S}_k)^-
\mid \psi_0 >= \biggr(1-\frac{1}{2}\frac{N-2}{N-1}\biggr) \mid \psi_1 >,
\end{equation}
thus
\begin{equation}
\label{N-eqq1n}
 -i \frac{2}{N}\sum_{j<k}^{N}[a(j)-a(k)] ({\vec S}_j \times {\vec S}_k)^-
\mid \psi_0 >= \mid \psi_1 >,
\end{equation}
on the other hand
\begin{equation}
\label{N-eqq1}
\biggr[\;Q_1^- =
-i \sum_{j<k}^{N} W_{jk}^{(1)}({\vec S}_j \times {\vec S}_k)^-\;\biggr]
\mid \psi_0>=\mid \psi_1>,
\end{equation}
by comparing the coefficients of $({\vec S}_j \times {\vec S}_k)^-$ of
the left-hand sides of Eqs.(\ref{N-eqq1n}) and (\ref{N-eqq1}) , it leads to
\begin{equation}
\label{N-newwjk1}
W_{jk}^{(1)}=\frac{2}{N}(a(j)-a(k)),\;\;
(j,k=1,2,\cdots,N)
\end{equation}
which is nothing but Eq. (\ref{N-eqwjk1}). Thus the physical picture for the
transformation from $\mid \psi_0>$ to $\mid \psi_1>$ is also clear.

\vspace{3mm}
\noindent
{\bf APPENDIX B: The Forms of Lowering Operators}
\vspace{3mm}

 Because $i ({\vec S}_j \times {\vec S}_k)^+=S_j^+ S_k^z - S_k^+S_j^z$,
one can calculate that

\begin{equation}
\label{N-eqsjk+2}
i ({\vec S}_j \times {\vec S}_k)^+ \mid \stackrel{j}{\uparrow}
\stackrel{k}{\downarrow} > = -\frac{1}{2}
\mid \stackrel{j}{\uparrow} \stackrel{k}{\uparrow} > ;
\;\;\;
i ({\vec S}_j \times {\vec S}_k)^+ \mid \stackrel{j}{\downarrow}
\stackrel{k}{\uparrow} > = \frac{1}{2}
\mid \stackrel{j}{\uparrow} \stackrel{k}{\uparrow} > .
\end{equation}
We set the lowering operator $Q_{0,1}^-$ is

\begin{equation}
\label{N-eqqr+3}
\biggr[\; Q_{0,1}^- = i \sum_{j<k}^{N} {W'_{jk}}^{(1)}
 ({\vec S}_j \times {\vec S}_k)^+ \;\biggr] \;
 \mid \psi_{1}>=\mid \psi_{0}>,
\;\; {W'_{jk}}^{(1)}=-{W'_{kj}}^{(1)},
\end{equation}
from which we obtain the equation

\begin{equation}
\label{N-eqwr'1}
\frac{1}{2} \sum_{m=1}^{N} \biggr[\;
 \biggr(-\sum_{j=1}^{m-1} {W'_{jm}}^{(1)}
 +\sum_{k=m+1}^{N} {W'_{mk}}^{(1)}\biggr)
 a(m) \;\biggr]=1,
\end{equation}
whose solutions are

\begin{equation}
\label{N-soluw'jk}
{W'_{j,k}}^{(1)}= \frac{2}{N(N-1)} \;\biggr[ \biggr(
\frac{\partial a(j)}{\partial j}\biggr)^{-1} \frac{\partial}{\partial j}-
  \biggr(\frac{\partial a(k)}{\partial k}
  \biggr)^{-1} \frac{\partial}{\partial k} \biggr].
\end{equation}

Because

\begin{equation}
i ({\vec S}_j \times {\vec S}_k)^+ \mid \stackrel{j}{\downarrow}
\stackrel{k}{\downarrow} > = \frac{1}{2}
(\mid \stackrel{j}{\downarrow} \stackrel{k}{\uparrow} > -
\mid \stackrel{j}{\uparrow} \stackrel{k}{\downarrow} > ),
\;\;\;
(S_j^++S_k^+)\mid \stackrel{j}{\downarrow} \stackrel{k}{\downarrow} >
=
(\mid \stackrel{j}{\downarrow} \stackrel{k}{\uparrow} > +
\mid \stackrel{j}{\uparrow} \stackrel{k}{\downarrow} > ),
\end{equation}
so that

$$
\mid \stackrel{j}{\downarrow} \stackrel{k}{\uparrow} >
=
\biggr[i ({\vec S}_j \times {\vec S}_k)^+   +
\frac{1}{2}(S_j^++S_k^+)\biggr]
\mid \stackrel{j}{\downarrow} \stackrel{k}{\downarrow} >,
$$
\begin{equation}
\label{N-spin+}
\mid \stackrel{j}{\uparrow} \stackrel{k}{\downarrow} >
=
\biggr[-i ({\vec S}_j \times {\vec S}_k)^+   +
\frac{1}{2}(S_j^++S_k^+)\biggr]
\mid \stackrel{j}{\downarrow} \stackrel{k}{\downarrow} >.
\end{equation}
Therefore, based on Eq. (\ref{N-spin+}) there would be some differences in
achieving lowering operators $Q_{r-1,r}^-$ when $r>1$. In the position,
we set

\begin{equation}
\label{N-eqqr3}
\biggr[\; Q_{r-1,r}^- = i \sum_{j<k}^{N} {W'_{jk}}^{(r)}
 ({\vec S}_j \times {\vec S}_k)^+
 + \sum_{j=1}^{N}
 \lambda_j^{(r)} S_j^+ \;\biggr]
 \mid \psi_{r}>=\mid \psi_{r-1}>,
\; {W'_{jk}}^{(r)}=-{W'_{kj}}^{(r)},
\end{equation}
 which is different from the raising operator by a translation term
 $\sum_{j=1}^{N} \lambda_j^{(r)} S_j^+$.

 For instance, when $r=2$, since
$$
  \mid \psi_2 >=\sum_{m_1<m_2} a(m_1,m_2) \mid \cdots
  \stackrel{m_1}{\downarrow} \cdots \stackrel{m_2}{\downarrow} \cdots  >,
$$
$$
  \mid \psi_1 >=\frac{1}{N-1} \sum_{m_1<m_2}[\; a(m_1) \mid \cdots
  \stackrel{m_1}{\downarrow} \cdots \stackrel{m_2}{\uparrow} \cdots  >+
  a(m_2) \mid \cdots
  \stackrel{m_1}{\uparrow} \cdots \stackrel{m_2}{\downarrow} \cdots  >\;],
$$
owing to Eq. (\ref{N-spin+}), one finds the transformation
$$
a(m_1,m_2) \mid \cdots
  \stackrel{m_1}{\downarrow} \cdots \stackrel{m_2}{\downarrow} \cdots  >
  \Longrightarrow
a(m_1) \mid \cdots
  \stackrel{m_1}{\downarrow} \cdots \stackrel{m_2}{\uparrow} \cdots  >
$$
can be realized by the operator:
\begin{equation}
 T_{(m_1,m_2) \rightarrow m_1}^-= \frac{a(m_1)}{a(m_1,m_2)}
      \biggr[\;
     i ({\vec S}_{m_1} \times {\vec S}_{m_2})^+  +
     \frac{1}{2}(S_{m_1}^+ + S_{m_2}^+) \;\biggr],
\end{equation}
and the transformation
$$
a(m_1,m_2) \mid \cdots
  \stackrel{m_1}{\downarrow} \cdots \stackrel{m_2}{\downarrow} \cdots  >
  \Longrightarrow
a(m_2) \mid \cdots
  \stackrel{m_1}{\uparrow} \cdots \stackrel{m_2}{\downarrow} \cdots  >
$$
is realized by the operator:
\begin{equation}
 T_{(m_1,m_2) \rightarrow m_2}^-= \frac{a(m_2)}{a(m_1,m_2)}
      \biggr[\;
    -i ({\vec S}_{m_1} \times {\vec S}_{m_2})^+  +
    \frac{1}{2}(S_{m_1}^+ + S_{m_2}^+) \;\biggr].
\end{equation}
Consequently
\begin{equation}
 [\; T_{(m_1,m_2) \rightarrow m_1}^-  +
     T_{(m_1,m_2) \rightarrow m_2}^- \;]
     a(m_1,m_2)\phi (m_1,m_2)=
     a(m_1)\phi (m_1) + a(m_2)\phi (m_2).
\end{equation}
Define
\begin{equation}
 F'_{(m_1,m_2)}=  [\; T_{(m_1,m_2) \rightarrow m_1}^-  +
     T_{(m_1,m_2) \rightarrow m_2}^- \;]
      a(m_1,m_2)
    \biggr(\frac{\partial^2 a(m_1, m_2)}{\partial m_1 \partial m_2 }
    \biggr)^{-1}\frac{\partial}{\partial m_1}\frac{\partial}{\partial m_2},
\end{equation}
so that
$$
\biggr(\sum_{m_1<m_2}^{N}F_{(m_1,m_2)}\biggr) \mid \psi_2 >  =
       (N-1) \mid \psi_1 >,
$$
on the other hand,
$$
 Q_{1,2}^- = i \sum_{j<k}^{N} {W'_{jk}}^{(2)}
 ({\vec S}_j \times {\vec S}_k)^+ + \sum_{j=1}^{N} \lambda_j^{(2)} S_j^+,
$$
thus
\begin{equation}
\label{N-eqq12f}
Q_{1,2}^- = \frac{1}{N-1}\sum_{m_1<m_2}^{N}F'_{(m_1,m_2)},
\end{equation}
by making comparison the coefficients of $({\vec S}_j \times {\vec S}_k)^+$
and $S_j^+$ of the both side of Eq. (\ref{N-eqq12f}), it yields
$$
{W'_{m_1,m_2}}^{(2)}=  \frac{1}{N-1} [a(m_1)-a(m_2)] \biggr(
\frac{\partial^2 a(m_1, m_2)}{\partial m_1 \partial m_2 }\biggr)^{-1}
\frac{\partial}{\partial m_1}\frac{\partial}{\partial m_2},
 \;\; (m_1< m_2),
$$
\begin{equation}
\label{N-soluw2'}
\lambda_j^{(2)} = \frac{1}{N-1}
 \sum_{m \neq j}^{N} [a(m)+a(j)]
     \biggr(\frac{\partial^2 A(m, j)}{\partial m \partial j}\biggr)^{-1}
\frac{\partial}{\partial m}\frac{\partial}{\partial j}.
\end{equation}

Lowering operators for $r \geq 3$ can be obtained in the same way, they are

$$
{W'_{jk}}^{(r)}=  \frac{1}{N-r+1}
\sum_{l_1,\cdots,l_{r-2}\neq j,k}
$$
$$
[A(j,l_1,\cdots,l_{r-2})-A(k,l_1,\cdots,l_{r-2})]
\biggr(\frac{\partial^r A(j,k,l_1,\cdots,l_{r-2})}
 { {\partial l_1}{\partial l_2}  \cdots {\partial l_{r-2}}
 {\partial j} {\partial k}}\biggr)^{-1}
 \frac{\partial}{\partial l_1} \frac{\partial}{\partial l_2}
 \cdots \frac{\partial}{\partial l_{r-2}}
  \frac{\partial}{\partial j}\frac{\partial}{\partial k}
$$
$$
\lambda_j^{(r)} = \frac{1}{N-r+1}
 \biggr\{ \sum_{m_1,\cdots,m_{r-1}\neq j}
 \biggr[\; \sum\limits_{(l_1,\cdots,l_{r-2})\in (m_1,\cdots,m_{r-1})}
  A(j,l_1,\cdots,l_{r-2})
$$
\begin{equation}
\label{soluw2''}
 + A(m_1,m_2,\cdots,m_{r-1}) \;\biggr]
\biggr(\frac{\partial^r A(j,m_1,m_2 \cdots,m_{r-1})}
 { {\partial m_1}{\partial m_2}  \cdots {\partial m_{r-1}}
 {\partial j} }\biggr)^{-1}
 \frac{\partial}{\partial m_1} \frac{\partial}{\partial m_2}
 \cdots \frac{\partial}{\partial m_{r-1}}
  \frac{\partial}{\partial j} \biggr\}.
\end{equation}

Like $Q_{r,r-1}^+$, when acts on $\mid \psi_{r}>$, $Q_{r-1,r}^-$ can be
simplified to ${\cal Q}_{r-1,r}^-= \sum\limits_{j=1}^{N} {\alpha'_j}^{(r)}
S_j^- $, with
\begin{equation}
\label{alph}
 {\alpha'_j}^{(r)}= \frac{1}{2} \sum_{k\neq j}^{N} {W'_{j,k}}^{(r)}
 +  \lambda_j^{(r)} .
\end{equation}
Consequently, the lowering operators $Q_{r-1,r}^-$ or
${\cal Q}_{r-1,r}^-$ are  also found.
In particular, ${\cal Q}_{1,0}^+ = \sum\limits_{m=1}^{N}a(m)S_m^-$,
${\cal Q}_{0,1}^-$ can be simplified to a more simple form
${\cal Q}_{0,1}^- = \sum\limits_{m=1}^{N}a^{-1}(m)S_m^+$ when it acts on
$\mid \psi_1>$. These two operators are mutually adjoint.
However, for general $r \geq 2$, the Hermitian properties for
${\cal Q}_{r,r-1}^+$ and ${\cal Q}_{r-1,r}^-$ are not held.

\end{document}